\documentclass[12pt]{iopart}
\usepackage{graphicx}
\usepackage{color}

\newcommand{\dfn}{\stackrel{\triangle}{=}}

\newcommand {\bh} {\mbox{\boldmath $h$}}

\newcommand {\bx} {\mbox{\boldmath $x$}}

\newcommand {\bH} {\mbox{\boldmath $H$}}

\newcommand{\calE}{{\cal E}}

\newcommand{\calH}{{\cal H}}
\newcommand{\calI}{{\cal I}}

\newcommand{\calS}{{\cal S}}

\newcommand{\calU}{{\cal U}}
\newcommand{\calV}{{\cal V}}
\newcommand{\calW}{{\cal W}}
\newcommand{\calX}{{\cal X}}

\begin{document}

\title[Sequence Complexity and Work Extraction]{Sequence Complexity
and Work Extraction}

\author{Neri Merhav}

\address{Department of Electrical Engineering, Technion, Haifa 32000,
Israel.\\ E--mail: merhav@ee.technion.ac.il}

\begin{abstract}
We consider a simplified version of a solvable model by Mandal and Jarzynski,
which constructively demonstrates the interplay between work extraction and
the increase of the Shannon entropy of an 
information reservoir which is in contact with a
physical system. We extend Mandal and Jarzynski's main findings in several directions:
First, we allow sequences of correlated bits rather than just independent
bits. Secondly, at least for the case of binary information, 
we show that, in fact, the Shannon entropy is only one measure of complexity of
the information that must increase in order for work to be extracted. The
extracted work can also be upper bounded in terms of the increase in other
quantities that measure complexity, like the predictability of future bits from past ones.
Third, we provide an extension to the case of non--binary
information (i.e., a larger alphabet), and finally, we extend the scope to the case
where the incoming bits (before the interaction) form an individual
sequence, rather than a random one. In this case, the entropy before the
interaction can be replaced by the Lempel--Ziv (LZ) complexity of the incoming
sequence, a fact that gives rise to an entropic meaning of the LZ complexity,
not only in information theory, but also in physics.
\end{abstract}

\maketitle

\section{Introduction}

Information processing and the role that it plays in thermodynamics is a
very well--known concept
that dates back to the second half of the nineteenth century, 
namely, to James Clerk Maxwell and his famous gedanken experiment,
known as Maxwell's demon \cite{maxwelldemon}. The Maxwell demon experiment shows that
an intelligent agent, with access to measurements of velocities and positions of particles
in a gas, is able to separate speedy particles from the slower ones, thereby
creating a temperature difference without injecting energy into the system,
which is seemingly in conflict with the second law of thermodynamics.
Several decades later, Leo Szilard \cite{szilardeng} continued this line of
thought, and demonstrated the conversion of heat into work, using a model of a box that
contains a single particle. He showed that by measurement and control, one may
be able to extract work in a closed cycle of the system, which is again, in
apparent contradiction with to the second law. 

This suspected violation of the
second law has triggered a long--lasting controversy and many
other thought--provoking gedanken experiments that have eventually furnished the basis
for a rather large of volume of theoretical work concerning the role and the
implications of information processing in thermodynamics. 
A non--exhaustive
list of recent works on the modern approach of incorporating informational
ingredients in physical systems includes 
\cite{BS13}, \cite{BS14}, \cite{CGQ15}, \cite{DJ13}, \cite{EV11}, \cite{GDC13},
\cite{HA14}, \cite{HE14}, \cite{HS14}, \cite{MJ12}, \cite{MQJ13},
\cite{PHS15}, \cite{SU12},
\cite{SU12a}, and \cite{SU13}.
In some of these
works, the informational resources are available by means of measurement and
feedback control (like in the Maxwell's demon and Szilard's engine) and
other works are about physical systems that include, in addition to the
traditional heat reservoir, also an {\it information
reservoir}, which interacts with the system, but without any energy
exchange. %This information reservoir can be, for example, a memory device, or
%a sequence of bits (recorded on a running tape), which interact serially with
%the system. 
The main common motive in these works is in extended versions of
the second law, where the expression of the 
entropy increase includes an extra entropic term that is associated with the information
exchange. These extended versions of the second law are, of course,
intimately related to Landauer's erasure principle \cite{Landauer61}.

Unlike earlier proposed thought experiments, that were mostly described in
generic terms and were not fully specified,
Mandal and Jarzynski \cite{MJ12} were the first to propose an explicit solvable model of a
concrete system that behaves in the spirit of the Maxwell demon. Specifically, they
described and analyzed a relatively simple autonomous system (based on a
six--state Markov jump process),
that when works as an engine, it converts thermal
fluctuations (heat) into mechanical work, while writing digital information onto a running
tape (in the role of an information reservoir), 
thereby increasing its Shannon entropy. It may also act as an eraser,
which implements the
opposite process of losing energy while erasing information, that is,
decreasing the entropy.
Several variations on this
model, based on similar ideas, were offered in some
subsequent works, e.g., \cite{BS13}, \cite{BS14}, \cite{CGQ15}, and \cite{MQJ13}.

In this paper, we consider a simplified version\footnote{Instead of the
six--state Markov process of \cite{MJ12}, we use a two--state process,
which is easier to analyze.}
of Mandal and Jarzynski's model \cite{MJ12} 
and we focus on extensions of their findings in several directions.
\begin{enumerate}
\item Allowing sequences of correlated bits rather than just independent
bits. 
\item At least for the case of binary information,
it is shown that, in fact, the Shannon entropy is only one measure of complexity
of the information that must increase in order for work to be extracted. The
extracted work can also be upper bounded in terms of the increase in other
quantities that measure complexity, like the predictability of future bits
from past ones.
\item An extension is offered for the case of non--binary
information (i.e., digital information with 
a larger alphabet). 
\item Extension of the scope to the
case where the incoming bits (before the interaction) form an individual
sequence, namely, a deterministic sequence rather than a random one. 
\end{enumerate}
In the last item above, instead of the term of information entropy before the
interaction, we have the Lempel--Ziv (LZ) complexity \cite{ZL78} of the incoming
sequence, a fact that gives rise to an entropic meaning of the LZ complexity,
not only in information theory, but also in physics. 

We believe that similar
extensions can be offered also for the other variations of this model, that
appear in \cite{BS13}, \cite{BS14}, \cite{CGQ15}, and \cite{MQJ13}, as
mentioned.

\section{Notation Conventions}

Throughout the paper, random variables will be denoted by capital
letters, specific values they may take will be denoted by the
corresponding lower case letters, and their alphabets
will be denoted by calligraphic letters. Random
vectors, their realizations and their alphabets will be denoted,
respectively, by capital letters, the corresponding lower case letters,
and the corresponding calligraphic letters,
all superscripted by their dimension.
For example, the random vector $X^n=(X_1,\ldots,X_n)$, ($n$ --
positive integer) may take a specific vector value $x^n=(x_1,\ldots,x_n)$
in $\calX^n$, which is the $n$--th order Cartesian power of $\calX$,
the alphabet of each component of this vector.
The probability of an event $\calE$ will be denoted by $P[\calE]$.
The indicator function of an event $\calE$ will be denoted by $\calI[\calE]$.

The Shannon entropy of a discrete random variable $X$ will be denoted\footnote{Following the
customary notation conventions in information theory, $H(X)$ should not be
understood as a function $H$ of the random outcome of $X$, but as a functional
of the probability distribution of $X$.} by $H(X)$, that is,
\begin{equation}
\label{entropydef}
H(X)=-\sum_{x\in\calX}P(x)\ln P(x),
\end{equation}
where $\{P(x),~x\in\calX\}$ is the probability distribution of $X$. 
When we wish to emphasize the dependence of the entropy on the underlying
distribution $P$, we denote it by $\calH(P)$. The binary entropy function will
be defined as
\begin{equation} 
h(p)=-p\ln p-(1-p)\ln(1-p),~~~~0\le p\le 1.
\end{equation} 
Similarly, for a discrete random vector $X^n=(X_1,\ldots,X_n)$, the joint entropy
is denoted by $H(X^n)$ (or by $H(X_1,\ldots,X_n)$), and defined as
\begin{equation}
\label{jointentropydef}
H(X^n)=-\sum_{x^n\in\calX^n}P(x^n)\ln P(x^n).
\end{equation}
The conditional entropy of a generic random variable $U$ over a discrete
alphabet $\calU$,
given another generic random variable
$V\in\calV$, is defined as
\begin{equation}
\label{condentropydef1}
H(U|V)=-\sum_{u\in\calU}\sum_{v\in\calV}P(u,v)\ln P(u|v), 
\end{equation}
which should not be confused with the conditional entropy given a {\it specific
realization} of $V$, i.e.,
\begin{equation}
\label{condentropydef1}
H(U|V=v)=-\sum_{u\in\calU}P(u|v)\ln P(u|v). 
\end{equation}
The mutual information between $U$ and $V$ is
\begin{eqnarray}
I(U;V)&=&H(U)-H(U|V)\nonumber\\
&=&H(V)-H(V|U)\nonumber\\
&=&H(U)+H(V)-H(U,V),
\end{eqnarray}
where it should be kept in mind that in all three definitions, $U$ and $V$ can
themselves be
random vectors. The Kullback--Leibler divergence (a.k.a.\ relative entropy or
cross-entropy) 
between two distributions $P$ and $Q$ on
the same alphabet $\calX$, is defined as
\begin{equation}
D(P\|Q)=\sum_{x\in\calX}P(x)\ln\frac{P(x)}{Q(x)}.
\end{equation}
%For two
%positive sequences $a_n$ and $b_n$, the notation $a_n\exe b_n$ will
%stand for asymptotic equivalence  in the exponential scale, that is,
%$\lim_{n\to\infty}\frac{1}{n}\log \frac{a_n}{b_n}=0$. Similarly,
%$a_n\lexe b_n$ means that
%$\limsup_{n\to\infty}\frac{1}{n}\log \frac{a_n}{b_n}\le 0$, and so on.
%The indicator function
%of an event $\calE$ will be denoted by $\calI\{E\}$. The notation $[x]_+$
%will stand for $\max\{0,x\}$.
%The empirical distribution of a sequence $\bx\in\calX^n$, which will be
%denoted by $\hat{P}$, is the set of relative frequencies
%$\hat{P}_{\bx}(x)$
%of each symbol $x\in\calX$ in $\bx$.

\section{Setup Description, Preliminaries and Objectives}

Consider the system depicted in the Fig.\ 1, which is a simplified
version of the one in \cite{MJ12}.

\begin{figure}[ht]
\hspace*{3cm}
\begin{picture}(0,0)%
\includegraphics{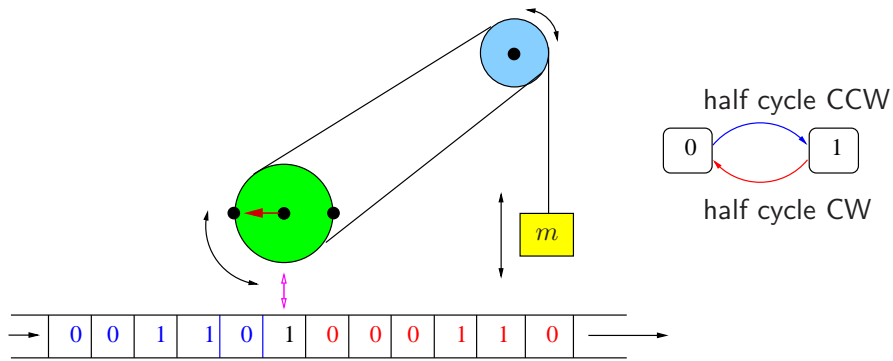}%
\end{picture}%
\setlength{\unitlength}{3947sp}%
\begingroup\makeatletter\ifx\SetFigFont\undefined%
\gdef\SetFigFont#1#2#3#4#5{%
  \reset@font\fontsize{#1}{#2pt}%
  \fontfamily{#3}\fontseries{#4}\fontshape{#5}%
  \selectfont}%
\fi\endgroup%
\begin{picture}(5959,2213)(145,-1541)
\put(4520,-654){\makebox(0,0)[lb]{\smash{{\SetFigFont{11}{13.2}{\sfdefault}{\mddefault}{\updefault}{half cycle CW}%
}}}}
\put(4520, 37){\makebox(0,0)[lb]{\smash{{\SetFigFont{11}{13.2}{\sfdefault}{\mddefault}{\updefault}{half cycle CCW}%
}}}}
\put(3467,-787){\makebox(0,0)[lb]{\smash{{\SetFigFont{10}{12.0}{\rmdefault}{\mddefault}{\itdefault}{$m$}%
}}}}
\end{picture}%
\caption{A system that interacts with a sequence of bits recorded on a running
tape.}
\end{figure}
\noindent

A device that consists of a wheel that is loaded (via a another wheel with
transmission) by a mass $m$, interacts with a running tape that
bears digital information in the form of a series of incoming bits, denoted $x_1,x_2,\ldots$,
$x_i\in\{0,1\}$, $i=1,2,\ldots$. The device also interacts thermally with a
heat bath at temperature $T$ (not shown in Fig.\ 1) in the form of heat exchange, but there is no
energy exchange with the tape.
During each time interval of $\tau$ seconds,
$i\tau\le t < (i+1)\tau$ ($i$ -- positive integer), 
the device interacts with the $i$--th bit, $x_i$, in
the following manner: If $x_i=0$, then the initial state of the
composite system (device plus bit) is `0' and then, due to random thermal
fluctuations, the wheel may spontaneously rotate, say, half a cycle
counter--clockwise (CCW) at a random time, thereby changing the state of the
system to `1'
and thus causing the mass to be lifted by $\Delta$ (which is half the
circumference of
the bigger wheel in Fig.\ 1).
Then, at a later random time, it may rotate clockwise (CW), changing the
state back to `0', and causing the mass to descend back by $\Delta$, etc. 
The net change in the height of the mass, 
during this interval, depends, of course, only on the parity of
the number of state transitions during this interval.
At the end of this time interval, namely, at
time $t=(i+1)\tau-0$, the current state is recorded on the tape as the
outgoing bit,
denoted by $y_i$. Note that if $x_i=0$ and $y_i=1$, then the net work done by the
device, during this time interval, is $\Delta W_i=mg\Delta$; otherwise
$\Delta W_i=0$.
Similarly, if the incoming bit is $x_i=1$, then the initial state is `1` and
then the first state transition (if any) is
associated with a CW rotation. By the same reasoning 
as before, at the end of the time interval,
if $y_i=0$, then the net work done by the device, during this interval, is $\Delta W_i=-mg\Delta$,
otherwise, it is $\Delta W_i=0$. 
Thus, in general, the work done during the $i$--th interval is $\Delta
W_i=mg\Delta\cdot(y_i-x_i)$.
Next, a new interval begins and it becomes 
the turn of bit $x_{i+1}$ to interact with
the device for $\tau$ seconds, and so on. 
It should be emphasized that this transition from the former outgoing bit
$y_i$ to a new incoming bit $x_{i+1}$ is not
accompanied by any energy exchange between the tape and the system
(the wheel does not move in response to this transition). 
This new bit just determines which
direction of rotation is enabled in which one is disabled.

The above described 
mechanism of back and forth transitions 
(with their associated rotations)
within each interval
is modelled as a two--state Markov jump process
with transition rates 
$\lambda_{0\to 1}$ and $\lambda_{1\to 0}$, related by
\begin{equation}
\lambda_{0\to 1}=\lambda_{1\to 0}e^{-mg\Delta/kT},
\end{equation}
giving rise to an equilibrium (Boltzmann) distribution
\begin{equation}
P_{\mbox{\tiny eq}}[0]=\frac{1}{1+e^{-mg\Delta/kT}};~~~
P_{\mbox{\tiny eq}}[1]=\frac{e^{-mg\Delta/kT}}{1+e^{-mg\Delta/kT}},
\end{equation}
which manifests the fact that state `1' is more energetic than state `0', the
energy difference being $\Delta E=mg\Delta$. At each interval, the 
temporal evolution
of the probability of state `1' is according to
the master equation:
\begin{equation}
\frac{\mbox{d} P_t[1]}{\mbox{d}t}=\lambda_{0\to 1}-\lambda P_t[1]
\end{equation}
where $\lambda\dfn\lambda_{0\to 1}+\lambda_{1\to 0}$. This simple first
order differential equation is readily solved by 
\begin{eqnarray}
P_t[1]&=&\frac{\lambda_{0\to 1}}{\lambda}+\left(P_0[1]-\frac{\lambda_{0\to
1}}{\lambda}\right)\cdot e^{-\lambda t}\nonumber\\
&=&P_{\mbox{\tiny eq}}[1]+\left(P_0[1]-P_{\mbox{\tiny eq}}[1]\right)\cdot
e^{-\lambda t},
\end{eqnarray}
and, of course, $P_t[0]$ complements to unity.
It is therefore readily seen that the
mechanism that transforms the sequence of incoming bits,
$x_1,x_2,\ldots$, into a sequence of outgoing bits, $y_1,y_2,\ldots$, 
is simply a
binary--input, binary--output 
discrete memoryless channel\footnote{A memoryless channel 
is characterized by the assumption that the conditional probability of $y^n$
given $x^n$ is given by the product of conditional probabilities of $y_i$
given $x_i$, $i=1,2,\ldots,n$.}
(DMC) $Q=[Q_{x\to y},~x,y\in\{0,1\}]$,
whose transition probabilities are given by
\begin{eqnarray}
Q_{0\to 0}&=&1-Q_{0\to 1}=P_{\mbox{\tiny eq}}[0]+P_{\mbox{\tiny eq}}[1]\cdot 
e^{-\lambda\tau}\\
Q_{1\to 1}&=&1-Q_{1\to 0}=P_{\mbox{\tiny eq}}[1]+P_{\mbox{\tiny eq}}[0]\cdot 
e^{-\lambda\tau}
\end{eqnarray}
The expected work done by the device after $n$ cycles is given by
\begin{eqnarray}
\left<W_n\right>&=&mg\Delta\cdot\left<\sum_{i=1}^n[Y_i-X_i]\right>\nonumber\\
&=&mg\Delta\cdot
\sum_{i=1}^n(P[Y_i=1]-P[X_i=1])\nonumber\\
&=&kTf\cdot
\sum_{i=1}^n(P[Y_i=1]-P[X_i=1]),
\end{eqnarray}
where $f\equiv mg\Delta/kT$.
Now, from the above derived time evolution of the
state distribution within an interval of duration $\tau$, 
one easily finds that
\begin{equation}
P[Y_i=1]=P_{\mbox{\tiny eq}}[1]+\left(P[X_i=1]-P_{\mbox{\tiny
eq}}[1]\right)\cdot e^{-\lambda
\tau},
\end{equation}
which means a monotonic
change, starting from $P[X_i=1]$ and ending at $P_{\mbox{\tiny eq}}[1]$.
In other words, $P[Y_i=1]$ is always between $P(X_i=1)$ and 
$P_{\mbox{\tiny eq}}[1]$.

We next focus on the informational (Shannon) entropy production, namely, the difference
between the entropy of the outgoing bit--stream $\{Y_i\}$ and the entropy of
the incoming bit--stream $\{X_i\}$.
By the concavity of binary entropy function, $h(\cdot)$, it is easily seen that for every
$s,t\in[0,1]$:
\begin{equation}
h(s)\le h(t)+(s-t)\cdot h'(t)\equiv h(t)+(s-t)\ln\frac{1-t}{t}.
\end{equation}
Thus, setting $s=P[X_i=1]$ and $t=P[Y_i=1]$, we get
\begin{eqnarray}
H(X_i)&\equiv&h(P[X_i=1])\nonumber\\
&\le&h(P[Y_i=1])+(P[X_i=1]-P[Y_i=1])\ln\frac{1-P[Y_i=1]}{P[Y_i=1]}.
\end{eqnarray}
or equivalently,
\begin{equation}
(P[Y_i=1]-P[X_i=1])\ln\frac{1-P[Y_i=1]}{P[Y_i=1]}\le H(Y_i)-H(X_i).
\end{equation}
Now, if $P[Y_i=1]\ge P[X_i=1]$, then $P_{\mbox{\tiny eq}}[1]\ge P[Y_i=1]\ge
P[X_i=1]$, and then 
\begin{eqnarray}
(P[Y_i=1]-P[X_i=1])\cdot f&=&
(P[Y_i=1]-P[X_i=1])\ln\frac{1-P_{\mbox{\tiny eq}}[1]}{P_{\mbox{\tiny
eq}}[1]}\nonumber\\
&\le&(P[Y_i=1]-P[X_i=1])\ln\frac{1-P[Y_i=1]}{P[Y_i=1]}\nonumber\\
&\le&H(Y_i)-H(X_i).
\end{eqnarray}
Similarly, if $P[Y_i=1]\le P[X_i=1]$, then $P_{\mbox{\tiny eq}}[1]\le
P[Y_i=1]\le
P[X_i=1]$, and then again,
\begin{equation}
(P[Y_i=1]-P[X_i=1])\cdot f\le H(Y_i)-H(X_i)
\end{equation}
since the terms $f$ and $\ln\{(1-P[Y_i=1])/P[Y_i=1]\}$ are multiplied by 
$(P[Y_i=1]-P[X_i=1])$, which is now non--positive. Thus, in both cases, the
last inequality holds, and so, as is actually shown in \cite{MJ12}
\begin{equation}
\label{basicineq}
\left<\Delta W_i\right>=kTf\cdot(P[Y_i=1]-P[X_i=1])\le kT[H(Y_i)-H(X_i)].
\end{equation}
Summing from $i=1$ to $n$, the left--hand side of (\ref{basicineq}) gives 
\begin{equation}
\label{nrounds}
\left<W_n\right>=kTf\cdot\sum_{i=1}^n(P[Y_i=1]-P[X_i=1])\le kT\sum_{i=1}^n[H(Y_i)-H(X_i)],
\end{equation}
where left--hand--side (l.h.s.) is the total average total work
after $n$ cycles.
The exact total average work
is given by
\begin{eqnarray}
\left<W_n\right>&=&kTf\cdot(1-e^{-\lambda\tau})
\left(nP_{\mbox{\tiny eq}}[1]-\sum_{i=1}^n
P[X_i=1]\right)\nonumber\\
&=&
kTf\cdot(1-e^{-\lambda\tau})\left(\sum_{i=1}^n
P[X_i=0]-nP_{\mbox{\tiny eq}}[0]\right),
\end{eqnarray}
which is obviously positive if and only if $\frac{1}{n}\sum_{i=1}^nP[X_i=0]>
P_{\mbox{\tiny eq}}[0]$.
If $\{X_i\}$ are i.i.d.\ (Bernoulli), as assumed in \cite{MJ12}
(as well as in subsequent follow--up papers mentioned earlier), then so are $\{Y_i\}$, and the
right--hand side (r.h.s.) of 
(\ref{nrounds}) agrees
with the total informational entropy production,
$kT\Delta H\dfn kT[H(Y^n)-H(X^n)]$.

As discussed in \cite{MJ12}, the inequality is saturated 
(in the sense that the ratio $f\cdot(P[Y_i=1]-P[X_i=1])/[H(Y_i)-H(X_i)]$ tends
to unity) when
$P[Y_i=1]$ is very close to $P[X_i=1]$ (which happens if either $\lambda\tau
\ll 1$ or if $P[X_i=1]$ is very close 
to $P_{\mbox{\tiny eq}}[1]$, to begin with), but then the amount of work
accumulated is very
small. To approach the entropy difference limit when this difference is
appreciably large,
one may iterate in small steps, namely, work with $\lambda\tau \ll 1$ and feed
$\{Y_i\}$ as an incoming bit--stream to another (identical, but independent) copy of 
the same device to generate, yet another bit--stream $\{Z_i\}$ with a further increased
entropy, etc. Alternatively, one may feed $\{Y_i\}$ back to the same system.
This way, with many repetitions of this process, the total work would be very close to 
$kT$ times the overall growth of the Shannon entropy. This idea is in the spirit of
quasi--static reversible processes in thermodynamics and statistical
mechanics.

As explained in the Introduction, we
extend these results in several directions:
\begin{enumerate}
\item Allowing the incoming bits, $X_1,X_2.\ldots,X_n$, 
to be correlated rather than just independent, identically distributed
(i.i.d.) bits.
In this case, the sum of entropy differences, $\sum_i[H(Y_i)-H(X_i)]$, at the r.h.s.\ of
(\ref{nrounds}) is different, in general, from the correct expression of the increase in the total Shannon
entropy, $H(Y^n)-H(X^n)$, which in turn takes the correlations among the bits 
into account. It
will be shown, nevertheless, that the correct expression associated with the entropy
increase, $kT[H(Y^n)-H(X^n)]$, is still an upper bound on the average work. 
This holds true for an arbitrary joint distribution of $(X_1,X_2,\ldots,X_n)$.
\item At least for the case of binary information,
it will be shown that an inequality like (\ref{basicineq}) (even in its
vector form) may hold even if the Shannon entropies on the r.h.s.\ are
replaced by generalized entropies, which may serve as alternative measures 
of information complexity, such as the average probability of error in predicting the next
bit $X_{i+1}$ from the bits seen thus far $X_1,X_2,\ldots,X_i$, $i=1,2,\ldots,n$.
\item We provide an extension of the above to the case of non--binary
information, i.e., $\{X_i\}$ and $\{Y_i\}$ take on values in a general finite
alphabet, whose size may be larger than 2. 
Under the general
alphabet size setting, however, item (ii) above is no longer claimed.
\item We extend the scope to the
case where the incoming bits $x_1,x_2,\ldots,x_n$ form an individual
sequence, namely, a deterministic sequence rather than a random one.
In this case, in the r.h.s.\ of (\ref{nrounds}), 
the analogue of the probabilistic input entropy $H(X^n)$ 
will be (for large $n$)
the Lempel--Ziv (LZ) complexity of the given sequence $x_1,x_2,\ldots,x_n$.
As for the output entropy ($Y^n$ is still a random vector), we will provide
computable bounds.
\end{enumerate}

\section{Correlated Input Bits}

Consider the case where the binary random vector
$(X_1,\ldots,X_n)$, of the first $n$ input bits, has a general
joint distribution, As said, in this case, the r.h.s.\ of eq.\ (\ref{nrounds}) 
is no longer associated with the correct overall change in the Shannon entropy,
$H(Y^n)-H(X^n)$. Nonetheless, our purpose, in this section, is to show that
the latter expression (times $kT$) continues to be an upper bound on the
expected work.

We proceed as follows.
Using the fact that channel $Q$ connecting $X^n$ and $Y^n$ is a DMC:
\begin{eqnarray}
\label{chain}
H(Y^n)-H(X^n)&=&\sum_{i=1}^n[H(Y_i|Y^{i-1})-H(X_i|X^{i-1})]\nonumber\\
&\ge&\sum_{i=1}^n[H(Y_i|X^{i-1},Y^{i-1})-H(X_i|X^{i-1})]\nonumber\\
&=&\sum_{i=1}^n[H(Y_i|X^{i-1})-H(X_i|X^{i-1})]\nonumber\\
&=&\sum_{i=1}^n\sum_{x^{i-1}}P(x^{i-1})[H(Y_i|X^{i-1}=x^{i-1})-\nonumber\\
& &H(X_i|X^{i-1}=x^{i-1})]\nonumber\\
&=&\sum_{i=1}^n\sum_{x^{i-1}}P(x^{i-1})\{h(P[Y_i=1|X^{i-1}=x^{i-1}])-\nonumber\\
& &h(P[X_i=1|X^{i-1}=x^{i-1}])\}\nonumber\\
&\ge&f\cdot
\sum_{i=1}^n\sum_{x^{i-1}}P(x^{i-1})(P[Y_i=1|X^{i-1}=x^{i-1}]-\nonumber\\
& &P[X_i=1|X^{i-1}=x^{i-1}])\nonumber\\
&=&f\cdot \sum_{i=1}^n(P[Y_i=1]-P[X_i=1])\nonumber\\
&=&\frac{\left<W_n\right>}{kT},
\end{eqnarray}
where the third line is due to the fact that 
$Y_i$ is statistically independent of $Y^{i-1}$ given $X^{i-1}$, and the
second inequality is again due to the concavity of $h(\cdot)$.\\

\noindent
{\bf Discussion.}
We have two upper bounds on the total work, $kT\sum_{i=1}^n[H(Y_i)-H(X_i)]$
and $kT[H(Y^n)-H(X^n)]$. As an upper bound, the former is always tighter, in
other words,
we argue (see Appendix A for the proof) that 
\begin{equation}
\label{entineq}
H(Y^n)-H(X^n)\ge \sum_{i=1}^n[H(Y_i)-H(X_i)],
\end{equation}
and so for the purpose of bounding the expected work, there is no point in looking
at higher order entropies of the incoming and outgoing processes.
However, from the physical point of view, the
inequality $\left<W_n\right>\le kT[H(Y^n)-H(X^n)]$ remains meaningful since the
difference $k[H(Y^n)-H(X^n)]-\left<W_n\right>/T$ has the natural meaning of the
total entropy production (of the combined system and
its environment) for the more general
case considered, i.e., where $\{X_i\}$ may be correlated. The
non--negativity of this difference is then a version of the
(generalized) second law of thermodynamics for systems that include
information reservoirs.
It follows from this discussion that if one has any
control on the incoming bit
sequence, then introducing
correlations among them is counter--productive in the sense that
it only enlarges the entropy
production without enlarging the extracted work (for a
given marginal probability assignment).
In other words, among all input vectors with a given
average marginal, $\bar{P}[x]=\frac{1}{n}\sum_{i=1}^nP[X_i=x]$,
the best one is an i.i.d.\ process (i.e., a Bernoulli process) with
a single--bit marginal given by $P[X_i=x]=\bar{P}[x]$ for all $i$.
In any other case,
there is an extra entropy production due to input correlations.

Note that if $X^n$ is a codeword from a rate--$R$ channel block code (with
equiprobable messages) for reliable communication
across the channel $Q$, namely, $H(X^n)=nR$ and $H(X^n|Y^n)$ is small by
Fano's inequality \cite[Section 2.10]{CT06}), then
\begin{eqnarray}
H(Y^n)-H(X^n)&\approx&H(Y^n|X^n)\nonumber\\
&=&\sum_{i=1}^n H(Y_i|X_i)=n[\bar{P}[0]h(Q_{0\to
0})+\bar{P}[1]h(Q_{1\to 1})].
\end{eqnarray}
In this case, as $H(Y^n)\approx n[R+\bar{P}[0]h(Q_{0\to
0})+\bar{P}[1]h(Q_{1\to 1})]$,
one can reliably recover from $Y^n$ both the incoming process $X^n$ and the entire
history of
of (net) movements of the wheel across the various intervals, so no
information is lost.

\section{Other Measures of Sequence Complexity}

Note that the only properties of the entropy function that were used
in Section 3 were: (i) concavity, and (ii) $h'(P_{\mbox{\tiny eq}}[1])=f$.
The second property does not pose any serious limitation because any concave function
can either be scaled or added with a linear term (both without harming the
concavity property), so that (ii) would hold. It follows then that
the Shannon entropy is not the only measure that describes
the increased complexity of information that must
accompany the extracted work. In other words, there are additional measures for
the amount extra randomness or the ``amount of information'' that must be
written in order to make the system convert heat to work.

We describe a generalized entropy function that is based on
a function $L_x(w)$, which is an arbitrary
function of $x\in\{0,1\}$ and a variable $w\in\calW$, that can be thought of
as a `loss' associated
with the choice of $w$ when the observation is $x$. We then define
a generalized entropy
function as the minimum achievable 
average loss associated with a binary random variable $X$, with $P[X=1]=1-P[X=0]=p$, that is
\begin{equation}
\bh(p)=\min_{w\in\calS}[(1-p)\cdot L_0(w)+p\cdot L_1(w)].
\end{equation}
Indeed, the binary Shannon entropy $h(p)$ 
is obtained as a special case for $L_0(w)=-\ln(1-w)$ and
$L_1(w)=-\ln w$, $\calW=[0,1]$, as the minimum is attained for $w^*=p$.
Since $\bh(p)$ is the minimum of affine functions of $p$, it is clearly
concave. Two additional examples of entropy--like functions are the following:
\begin{enumerate}
\item Let
$L_x(w)=\calI[w\neq x]$, $\calW=\{0,1\}$, measure the loss in
(possibly erroneous) `guessing' of $x$ by $w$. In this case,
$\bh(p)=\min\{p,1-p\}$. 
\item 
The squared--error loss function, $L_x(w)=(x-w)^2$, $\calW=[0,1]$, yields $\bh(p)=p(1-p)$.
\end{enumerate}
The extension of (\ref{basicineq}) now asserts 
that the average work extraction 
$\left<\Delta W_i\right>$, within a single cycle, cannot exceed
\begin{equation}
\frac{mg\Delta}{\bh'(P_{\mbox{\tiny eq}}[1])}\cdot \Delta\bh=
\frac{kTf}{\bh'(P_{\mbox{\tiny eq}}[1])}\cdot \Delta\bh,
\end{equation}
where $\Delta\bh=\bh(P[Y_i=1])-\bh(P[X_i=1])$ is the increase
in the (generalized) `complexity' in $Y_i$ relative to $X_i$, and where we
have assumed that $\bh(\cdot)$ is differentiable at $p=P_{\mbox{\tiny
eq}}[1]$. We will comment on the non--differentiable case shortly.

Denoting $\bH(X_i)=\bh(P[X_i=1])$
and $\bH(Y_i)=\bh(P[Y_i=1])$, we can generalize the above discussion
(including (\ref{chain}), provided that the first equality is considered a
definition) to correlated sequences of bits,
by introducing the definition
\begin{equation}
\bH(X_i|X^{i-1})=\sum_{x^{i-1}}P(x^{i-1})\bh(P[X_i=1|X^{i-1}=x^{i-1}])
\end{equation}
and similar definitions for the other generalized conditional entropies.
Considering the first example above,
$\bH(X_i|X^{i-1})$ designates the {\it
predictability} \cite{fmg92} of
$X_i$ given $X^{i-1}$, i.e., the minimum achievable probability of error in guessing
$X_i$ from $X^{i-1}$, which is
certainly a reasonable measure of complexity.
As for the second example above,
$\bH(X_i|X^{i-1})$ has the meaning of the minimum mean squared error
in estimating $X_i$ based on $X^{i-1}$. Here, $\bh'(P_{\mbox{\tiny
eq}}[1]))=\tanh(f/2)$. Thus, the factor $kTf/\bh'(P_{\mbox{\tiny
eq}}[1]))=kTf/\tanh(f/2)$, which is about $kTf=mg\Delta$ at very low
temperatures, and about $2kT$ at very high temperatures.

On a technical note,
observe that in general $\bh(\cdot)$ 
may not be differentiable at $P_{\mbox{\tiny
eq}}[1]$, but due to the concavity, 
there are always one--sided derivatives
$\bh_+'(P_{\mbox{\tiny eq}}[1])=\lim_{\delta\downarrow 0}[\bh(P_{\mbox{\tiny
eq}}[1]+\delta)-\bh(P_{\mbox{\tiny eq}}[1])]/\delta$ and
$\bh_-'(P_{\mbox{\tiny eq}}[1])=\lim_{\delta\uparrow 0}[\bh(P_{\mbox{\tiny
eq}}[1]+\delta)-\bh(P_{\mbox{\tiny eq}}[1])]/\delta$,
with $\bh_-'(P_{\mbox{\tiny eq}}[1])\geq
\bh_+'(P_{\mbox{\tiny eq}}[1])$.
We can always use either one. In case of a strict inequality, we
can choose the one that gives the tighter inequality,
namely, $\bh_-'(P_{\mbox{\tiny
eq}}[1])$ if
$\sum_i P[Y_i=1]\ge \sum_iP[X_i=1]$ and $\bh_+'(P_{\mbox{\tiny eq}}[1])$
otherwise.

Another class of generalized entropies obey the form $\bH(X)=
\left<S[1/P(X)]\right>$, where $S$ is am
arbitrary concave function (e.g., $S[u]=\ln u$ gives the
Shannon entropy), which is easily seen to be
concave functional of $P$. 
%Letting $\lambda_1\ge 0$ and $\lambda_2\ge 0$ sum
%to unity, and letting $P(x)=\lambda_1P_1(x)+\lambda_2P_2(x)$,
%\begin{eqnarray}
%\lambda_1\bH(X_1)+
%\lambda_2\bH(X_2)
%&=&\sum_x
%P(x)\left\{\frac{\lambda_1P_1(x)}{P(x)}Q\left[\frac{1}{P_1(x)}\right]+
%\frac{\lambda_2P_2(x)}{P(x)}Q\left[\frac{1}{P_2(x)}\right]\right\}\nonumber\\
%&\le&\sum_x
%P(x)Q\left[\frac{1}{P(x)}\right]=\bH(X).
%\end{eqnarray}
In the binary case considered
here, this would amount to $\bh(p)=pS[1/p]+(1-p)S[1/(1-p)]$.
The concavity property guarantees
that our earlier arguments hold for this kind of generalized entropy as well.
Similar comments apply to yet another class of generalized entropies,
$\bH(X)=\sum_x S[P(x)]$, where $S$ is again concave (e.g., $S[u]=-u\ln u$ gives the
Shannon entropy). 

This discussion sets the stage for a richer 
family of bounds on the extracted work, which
depend on various notions of sequence complexity. 
Provided that $\bh(\cdot)$ is differentiable at $P_{\mbox{\tiny
eq}}[1]$, these bounds are asymptotically met in the limit of infinitesimally small
differences between $P[Y_i=1]$ and $P[X_i=1]$, as discussed above
in the context of the ordinary entropy.
Nonetheless, among all generalized entropies we have discussed, 
only the Shannon entropy is
known to be
invariant under permutations, e.g., for $n=2$,
$H(X_1)+H(X_2|X_1)=H(X_2)+H(X_1|X_2)$, but
in general, it not true that $\bH(X_1)+\bH(X_2|X_1)=\bH(X_2)+\bH(X_1|X_2)$.
Also, it is not clear if and how any of the other entropy--like functionals continue to
serve in bounding the average work when the setup is extended to larger
alphabets (see Section 6 below).
These two points give rise to the special stature of the ordinary Shannon
entropy, which prevails in a deeper sense and in more general
situations.

\section{Non--Binary Sequences}

In this section, we extend the results from the case where $\{x_i\}$ and
$\{y_i\}$ taken on binary values to the case of a general finite alphabet of
size $K$. Correspondingly, in this case, the underlying system dymamics would
be associated with some Markov jump process having $K$ states.
Associated with each state $s$, there would a corresponding height increment,
$\Delta(s)$ (which may be positive or negative), relative to some reference
state, say $s_0$, with
$\Delta(s_0)\equiv 0$. Each state $s$ is then associated with energy,
$E(s)=mg\Delta(s)$,
and the transition rates of the underlying Markov process obey detailed balance accordingly.
For the given interaction interval of length $\tau$, let $P_t$ denote the
$K$-dimensional vector of state probabilities at time $t$, so that the initial
distribution vector $P_0$ corresponds to the probability distribution of the
incoming symbol, the final distribution $P_\tau$, corresponds to that of the
outgoing symbol, and $P_\infty=P_{\mbox{\tiny eq}}$. The Markovity of the
process implies 
that $D(P_t\|P_{\mbox{\tiny
eq}})$ is monotonically non--increasing\footnote{This monotonicity property
is, in fact, an extended version of the H-Theorem to the case where the
equilibrium distribution is not necessarily uniform. Informally speaking,
while the
H-Theorem is about the increase in (Shannon) entropy in an isolated system,
this monotonicity property of the divergence symbolizes the decrease in free energy more
generally.}
in $t$ (see, e.g., \cite{CT06}, \cite{p139}),
and so,
\begin{equation}
D(P_\tau\|P_{\mbox{\tiny eq}})\le
D(P_0\|P_{\mbox{\tiny eq}}),
\end{equation}
which after straightforward algebraic manipulation, becomes
\begin{equation}
\label{div}
\sum_s (P_{\tau}[s]-P_0[s])\ln\frac{1}{P_{\mbox{\tiny eq}}[s]}\le
\calH(P_\tau)-\calH(P_0).
\end{equation}
Since 
\begin{equation}
\label{p2w}
\ln\frac{1}{P_{\mbox{\tiny
eq}}[s]}=\ln Z+\frac{mg\Delta(s)}{kT},
\end{equation}
$Z=\sum_s \exp\{-mg\Delta(s)/kT\}$ being the partition function,
the l.h.s.\ of (\ref{div}) gives the average work per
cycle (in units of $kT$), and the r.h.s.\ is, of course, the entropy difference.
Note that we have assumed nothing about the structure of the underlying Markov process
except detailed balance.
This discussion easily extends to the case of correlated input symbols, as in
Section 4. 

\section{Individual Sequences and the LZ Complexity}

Finally, we extend the scope to the case where $x_1,x_2,\ldots$
is an individual sequence, namely, 
an arbitrary deterministic sequence, with no assumptions concerning the
mechanism that has generated it. The outgoing sequence is, of course, still
random due to the randomness of the state transitions.
In this setting, the LZ complexity of the incoming sequence will play
a pivotal role, and therefore, before moving on to the derivation for the
individual--sequence setting, we pause to
provide a brief background concerning the LZ complexity, which can be thought
of as an individual--sequence counterpart of entropy.

In 1978, Ziv and Lempel \cite{ZL78} invented their famous universal algorithm
for data compression, which has been considered a major breakthrough, both from
the theoretical aspects and the practical aspects of data compression.
For an given (individual) infinite sequence, $x_1,x_2,\ldots$, the LZ algorithm
achieves a compression ratio, which is asymptotically as good as that of the
best data compression algorithm that is implementable by a finite--state
machine. To the first order, the compression 
ratio achieved by the LZ algorithm, upon compressing the first $n$
symbols, $x^n=(x_1,x_2,\ldots,x_n)$, i.e., the LZ complexity of $x^n$,
is about 
\begin{equation}
\rho(x^n)=\frac{c(x^n)\log c(x^n)}{n},
\end{equation}
where $c(x^n)$ is the number of distinct {\it phrases} of $x^n$ obtained upon
applying the so called {\it incremental parsing procedure}. The
incremental parsing procedure works as follows. 
The sequence $x_1,x_2,\ldots,x_n$ is parsed
sequentially (from left to right), where each parsed
phrase is the shortest string that has not been encountered 
before as a parsed phrase, except perhaps the last phrase, which might be
incomplete. For example, the sequence 
$x^{17}=10001101110100010$ 
is parsed as
$1,0,00,11,01,110,10,001,0$. The first two phrases are obviously 
`1´ and `0´ as there is
no `history' yet. The next `0' has 
already been seen as a phrase, but the string `00'
has not yet been seen, so the
next phrase is `00'. Proceeding to the next bit, 
`1' has already appeared as a phrase, but `11' has
not, and so on. In this example then, $c(x^{17})=9$.
The idea of the LZ algorithm is to sequentially compress the sequence 
phrase--by--phrase,
where each phrase, say, of length $r$, is 
represented by a pointer to the location of the appearance of the first
$r-1$ symbols as a previous phrase (already decoded by the de-compressor),
plus an uncompressed representation of the $r$-th symbol of that phrase.
It is shown in \cite{ZL78} that if the LZ algorithm is applied to a random
vector $X^n$, that is sampled from a stationary and ergodic process, then
$\rho(X^n)$ converges with probability one to the entropy rate of the process,
$\bar{H}=\lim_{n\to\infty} H(X_n|X^{n-1})$.
In that sense, $\rho(x^n)$ can be thought of as an analogue of entropy in the
individual--sequence setting.

The general idea, in this section, is that,
in the context of the entropic upper bound on the extracted work,
the role of the input entropy,
$H(X^n)$, of the probabilistic case, will now be played by $\rho(x^n)$,
whereas $H(Y^n)$ will be upper bounded in terms of $\rho(x^n)$.
Thus, the concept of LZ complexity is not only analogous to
information--theoretic entropy, but in a way, it also plays an entropic role in the
physical sense.

Equipped with this background, we now move on to the derivation.
For simplicity, we consider the binary case, but everything can be extended to
the non--binary case, following the considerations
of Section 6.
Consider then an individual binary sequence $(x_1,x_2,\ldots,x_n)$ of incoming
bits. Let $\ell$ be a divisor of $n$ and chop the sequence into
$n/\ell$ non-overlapping blocks of length $\ell$,
$\bx_i=(x_{i\ell+1},x_{i\ell+2},\ldots,x_{i\ell+\ell})$,
$i=0,1,\ldots,n/\ell-1$. Consider now the empirical distribution of
$\ell$--blocks
\begin{equation}
\hat{P}(x^\ell)=\frac{\ell}{n}\sum_{i=0}^{n/\ell-1}\calI[\bx_i=x^\ell],~~~~x^\ell\in\{0,1\}^\ell
\end{equation}
Now, define 
\begin{equation}
\hat{P}[X_i=1]=\sum \hat{P}(x^\ell),~~~~i=1,2,\ldots,\ell
\end{equation}
where the summation is over all binary $\ell$--vectors $\{x^\ell\}$ whose
$i$--th coordinate is $1$.
The average work for a given $(x_1,x_2,\ldots,x_n)$ is
given by
\begin{eqnarray}
\left<W_n\right>&=&kTf\cdot\sum_{t=1}^n(\left<Y_t\right>-x_t)\nonumber\\
&=&kTf\cdot\frac{n}{\ell}\cdot\sum_{i=1}^\ell (P[Y_i=1]-\hat{P}[X_i=1])\nonumber\\
&\le&\frac{kTfn}{\ell}\cdot[\tilde{H}(Y^\ell)-\hat{H}(X^\ell)]
\end{eqnarray}
where $P[Y_i=1]=\hat{P}[X_i=1]Q_{1\to 1}+\hat{P}[X_i=0]Q_{0\to 1}$,
$\hat{H}(X^\ell)$ is the empirical entropy of
$\ell$--blocks associated with $x^n=(x_1,x_2,\ldots,x_n)$ and
$\tilde{H}(Y^\ell)$ is the output entropy of $\ell$--vectors that is induced
by the input assignment $\{\hat{P}(x^\ell)\}$ and $\ell$ uses of the memoryless 
channel $Q$. The last inequality is simply an application of the results of
Section 4 to the case where the joint distribution of $X^n$ is
$\hat{P}(\cdot)$.
This already gives some meaning to the notion of
entropy production in this case, where the
incoming bits
are deterministic. However,
the choice of the parameter $\ell$ (among the divisors of $n$)
appears to be somewhat arbitrary.
In the following, we further obtain another bound, which is asymptotically, 
independent of $\ell$. In this bound, $\hat{H}(X^\ell)$ will be
replaced by $\rho(x^n)$.
From \cite[eq.\ (21)]{p148}, we have the following lower bound on $\hat{H}(X^\ell)$ in terms of
its LZ complexity (setting the alphabet size to $2$ 
and passing logarithms to base $e$):
\begin{eqnarray}
\label{clogclb}
\frac{\hat{H}(X^\ell)}{\ell}&\ge&\rho(x^n)-
\frac{8\ell\ln 2}{(1-\epsilon_n)\log n}-
\frac{2\ell 4^{\ell}\ln 2}{n}-\frac{\ln 2}{\ell}\nonumber\\
&\equiv&\rho(x^n)-\delta(n,\ell),
\end{eqnarray}
where $\epsilon_n\to 0$. This inequality is a result of comparing the
compression ratio of a certain block code to a lower bound on the compression
performance of a general finite--state machine, which is essentially
$\rho(x^n)$. Of course,
$\lim_{\ell\to\infty}\lim_{n\to\infty}\delta(n,\ell)=0$.
Let $\delta_n$ denote the minimum of 
$\delta(n,\ell)$ over all $\{\ell\}$ that are divisors of $n$.

It remains to deal with the entropy of $Y^n$.
First, observe that the case of very large $\tau$ is obvious, because in this
case, $H(Y^n)=n\calH(P_{\mbox{\tiny eq}})$ as $\{Y_i\}$
is i.i.d.\ with marginal $P_{\mbox{\tiny eq}}$, regardless of $x^n$. 
Therefore, neglecting the term $\delta_n$, the upper bound on the extracted
work becomes
\begin{equation}
\left<W_n\right>\le kTn[\calH(P_{\mbox{\tiny eq}})-\rho(x^n)].
\end{equation}
For a general $\tau$, we proceed as follows. 
Given the binary--input, binary--output 
DMC $Q:X\to Y$, define the single--letter function 
\begin{equation}
U(z)=\max\{H(Y):~H(X)\ge z\}.
\end{equation}
The function $U(z)$ is concave
and monotonically decreasing. The monotonicity is obvious. As for the
concavity, indeed,
let $P_0$ and $P_1$ be the achievers of $U(z_0)$ and $U(z_1)$,
respectively. Then, for $0\le \lambda\le 1$, the entropy of 
$P_\lambda=(1-\lambda)P_0+\lambda P_1$ 
is never less than $(1-\lambda)z_0+\lambda z_1$, and so,
\begin{eqnarray}
U[(1-\lambda)z_0+\lambda z_1]&\ge& H(Y_\lambda)~~~~~~Y_\lambda~\mbox{being
induced by 
$P_\lambda$ and $Q$}\nonumber\\
&\ge&(1-\lambda) H(Y_0)+\lambda H(Y_1)\nonumber\\
&=&(1-\lambda)U(z_0)+\lambda U(z_1).
\end{eqnarray}
Note that if $H(X^\ell)\ge \ell z$, then a--fortiori,
$\sum_{i=1}^{\ell} H(X_i)\ge \ell z$, and so,
for the given DMC,
\begin{eqnarray}
H(Y^\ell)&\le&\sum_{i=1}^\ell H(Y_i)\nonumber\\
&\le&\sum_{i=1}^\ell U[H(X_i)]\nonumber\\
&\le&\ell\cdot U\left[\frac{1}{\ell}\sum_{i=1}^\ell H(X_i)\right]\nonumber\\
&\le&\ell\cdot U(z).
\end{eqnarray}
Applying this to the input distribution $\{\hat{P}(x^\ell)\}$ and the channel
$Q$, we have, by (\ref{clogclb}):
\begin{equation}
\tilde{H}(Y^\ell)\le \ell\cdot U\left[\rho(x^n)
-\delta_n\right],
\end{equation}
and so, defining the function 
\begin{equation}
V(z)\equiv U(z)-z,
\end{equation}
which is concave and
decreasing as well,
we get the following upper bound on $\left<W_n\right>$ in terms of the LZ
complexity of $x^n$:
\begin{equation}
\left<W_n\right>\le kTfn\cdot V\left[\rho(x^n)
-\delta_n\right].
\end{equation}
It tells us, among other things, that
the more $x^n$ is LZ--compressible, 
the more work extraction one can hope for.

This upper bound is tight 
in the sense that 
no other bound that depends on $x^n$
only via its LZ compressibility $\rho(x^n)$ can be tighter, 
because for a given value $\rho$ (in the range where the constraint
in the maximization defining $U(\rho)$ is attained with equality) of
the LZ compressibility, $\rho(x^n)$, there exist sequences with
LZ compressibility $\rho$ for which the bound $kTfnV(\rho)$ is essentially
attained. This is the case, for example, for 
most typical sequences of the memoryless
source $P^*$ that achieves $U(\rho)$. Tighter bounds can be obtained, of
course, if more
detailed information is given about the empirical statistics of $x^n$.

The important point about the function $U$ (and, of course, $V$) is that, in the jargon of
information theorists,
it is a single--letter function, that is, its calculation
requires merely optimization in the level of marginal distributions of a
single symbol, and not distributions associated with $\ell$--vectors.
In Appendix B, we provide an explicit expression of $U(z)$. 

\section*{Acknowledgement}

This research was supported by the Israel Science Foundation (ISF), grant no.\
412/12.

\section*{Appendix A}

\noindent
{\it Proof of Eq.\ (\ref{entineq}).}

The proof is by induction: For $n=1$, this is trivially true. Assume that it
is true for a given $n$. Then, by the memorylessness of the channel $Q$,
$Y^n\to X^n\to X_{n+1}\to Y_{n+1}$ is a Markov chain, and so, by the data
processing theorem \cite[Section 2.8]{CT06}
\begin{eqnarray}
H(X^n)+H(X_{n+1})-H(X^{n+1})&=&I(X_{n+1};X^n)\nonumber\\
&\ge& 
I(Y_{n+1};Y^n)\nonumber\\
&=&H(Y^n)+H(Y_{n+1})-H(Y^{n+1}),
\end{eqnarray}
which is equivalent to
\begin{equation}
H(Y^{n+1})-H(X^{n+1})\ge [H(Y^n)-H(X^n)]+[H(Y_{n+1})-H(X_{n+1})],
\end{equation}
and so,
\begin{equation}
H(Y^n)-H(X^n)\ge \sum_{i=1}^n[H(Y_i)-H(X_i)]
\end{equation}
implies
\begin{equation}
H(Y^{n+1})-H(X^{n+1})\ge \sum_{i=1}^{n+1}[H(Y_i)-H(X_i)],
\end{equation}
completing the proof.

\section*{Appendix B}

\noindent
{\it Deriving an Explicit Expression for $U(z)$.}

For the case of the
binary--input, binary--output channel at hand, let us denote
$\epsilon_0=Q_{0\to 0}$ and $\epsilon_1=Q_{1\to 0}$, and assume that
$\epsilon_0\ge\epsilon_1$ (otherwise, switch the roles of the inputs).
If the input assignment is $(p,1-p)$, then the output entropy is clearly
$h(p\epsilon_0+\bar{p}\epsilon_1)$ ($\bar{p}$ being $1-p$). The constraint
$h(p)\ge z$ is equivalent to the constraint $h^{-1}(z)\le p\le 1-h^{-1}(z)$,
where $h^{-1}(s)$ is the smaller of the two solutions $\{u\}$ to the equation
$h(u)=z$. Denoting
\begin{eqnarray}
\alpha_z&=&\epsilon_0h^{-1}(z)+\epsilon_1[1-h^{-1}(z)]\\
\beta_z&=&\epsilon_0[1-h^{-1}(z)]+\epsilon_1h^{-1}(z)
\end{eqnarray}
then $\epsilon_0\ge\epsilon_1$ implies $\beta_z\ge\alpha_z$, and then
\begin{eqnarray}
U(z)&=&\max\{h(q):~\alpha_z\le q\le\beta_z\}\nonumber\\
&=&\left\{\begin{array}{ll}
h(\beta_z) & \beta_z\le \frac{1}{2}\\
\ln 2 & \alpha_z\le \frac{1}{2} \le \beta_z\\
h(\alpha_z) & \alpha_z\ge \frac{1}{2}\end{array}\right.
\end{eqnarray}
The condition $\beta_z\le 1/2$ is satisfied always if $\epsilon_0\le 1/2$.
For $\epsilon_0> 1/2\ge \epsilon_1$, this condition is equivalent to
\begin{equation}
z\ge h\left(\frac{\epsilon_0-1/2}{\epsilon_0-\epsilon_1}\right)\equiv z^*
\end{equation}
Similarly, the condition
$\alpha_z\ge 1/2$ is satisfied always if $\epsilon_1\ge 1/2$.
For $\epsilon_1< 1/2\le \epsilon_0$, this condition is equivalent to
\begin{equation}
z\ge h\left(\frac{1/2-\epsilon_1}{\epsilon_0-\epsilon_1}\right)=z^*
\end{equation}
Thus, to summarize, $U(z)$ behaves as follows:
\begin{enumerate}
\item For $\epsilon_1 \ge 1/2$, $U(z)=h(\alpha_z)$ for all $z\in[0,1]$.
\item For $\epsilon_0 \le 1/2$, $U(z)=h(\beta_z)$ for all $z\in[0,1]$.
\item For $\epsilon_1 \le 1/2 \le \epsilon_0$ and $\epsilon_0+\epsilon_1 > 1$
$$U(z)=\left\{\begin{array}{ll}
\ln 2 & 0\le z\le z^*\\
h(\alpha_z) & z^*< z\le 1\end{array}\right.$$
\item For $\epsilon_1 \le 1/2 \le \epsilon_0$ and $\epsilon_0+\epsilon_1 < 1$
$$U(z)=\left\{\begin{array}{ll}
\ln 2 & 0\le z\le z^*\\
h(\beta_z) & z^*< z\le 1\end{array}\right.$$
\end{enumerate}
Note that for the binary symmetric channel ($\epsilon_0+\epsilon_1=1$), 
trivially $U(z)\equiv\ln 2$ for all $z\in[0,1]$.
Also, in all cases $U(1)=h[(\epsilon_0+\epsilon_1)/2]$.

\end{document}